\documentstyle[prl,aps,eqsecnum,epsfig]{revtex}
\tighten\begin{document}
\preprint{UCSBTH-96-23; CMU-HEP-96-10; DOR-ER/40682-121; PITT-96-2235; LPTHE-96-40}
\draft
\title{\bf PREHEATING IN FRW UNIVERSES}
\author{{\bf D. Boyanovsky$^{(a)}$, D. Cormier$^{(b)}$,  H.J. de Vega$^{(c)}$,
R. Holman$^{(b)}$, A. Singh$^{(d)}$ and M. Srednicki$^{(d)}$}}
\address
{{\it (a)  Department of Physics and Astronomy, University of
Pittsburgh, Pittsburgh, PA. 15260, U.S.A.} \\
{\it (b) Department of Physics, Carnegie Mellon University, Pittsburgh,
PA. 15213, U. S. A.} \\
{\it (c) Laboratoire de Physique Th\'eorique et Hautes Energies$^{[*]}$
Universit\'e Pierre et Marie Curie (Paris VI), 
Tour 16, 1er. \'etage, 4, Place Jussieu 75252 Paris, Cedex 05, France}\\
{\it (d) Department of Physics, University of
California, Santa Barbara, CA 93106, U. S. A.}}
\date{September 1996}
\maketitle
\begin{abstract}
The nonlinear time evolution of the quantum fields is studied in
the $O(N)$ model for large $N$  in a radiation dominated FRW universe,
with a view towards  the phenomenon 
of explosive particle production due to either spinodal instabilities or
parametric amplification, i.e. preheating. Quantum backreaction effects due to
the produced particles are included consistently within the large $N$
approximation. We find that preheating persists when the expansion is included,
although the amount of particle production is reduced compared to the values
found in Minkowski space. We also see that the behavior of the evolving zero
mode is very different from that in Minkowski space, though the late time
behavior in all cases is determined by a sum rule that implies the existence of
Goldstone bosons in the final state.
\end{abstract}

\section{Introduction}

A viable model of inflation must be able to reheat the Universe to high enough
temperatures to regain the tested features of the standard Big-Bang model, such
as nucleosynthesis. However, it has recently been noted\cite{branden,kls,us1}
that the theory of reheating as originally conceived\cite{origreheat} requires
revision due to abundant particle production originating from the coupling of
the zero mode of the inflaton field to other modes of this same field or modes
of a different field (see also refs.\cite{par,tkachev,son,symrest}).

Preheating is an inherently non-equilibrium process and as such, any field
theoretical treatment must take this into account. Preheating is also a
non-perturbative process; if $\lambda$ is either the self-coupling of the given
field or the coupling of this field to another, then, in typical scenarios for 
which the initial energy in the zero mode is ${\cal O}(1\slash \lambda)$,
the number of particles produced is typically ${\cal O}(1\slash \lambda)$.  
Thus any perturbative treatment will necessarily be incomplete.

In the Minkowski space analysis of refs.\cite{us1,us2}, these constraints were
dealt with by applying the so-called closed time path formalism (CTP)\cite{ctp}
of quantum field theory, which describes the non-equilibrium time evolution of
the density matrix describing the state of the field theory, to both the
Hartree approximation of a quartically self-coupled scalar theory as well as
the large $N$ approximation to the $O(N)$ vector
model\cite{largeN,losalamos}. These analyses clearly displayed the
dissipational effects of backreaction due to the produced particles as the zero
mode evolved towards its final equilibrium state. They also showed that the
zero mode can execute extremely interesting and unexpected behavior, depending
upon the couplings as well as the initial conditions\cite{par}.

In this work, we ask the question: does the inclusion of the Hubble expansion
affect the preheating process? Some partial answers to this question were given
in ref.\cite{kaiser,yoshimura}.

We consider the $O(N)$ model embedded in a {\em fixed} radiation dominated FRW
background spacetime. Again, we include the effects of backreaction to leading
order in $1 \slash N$ and calculate the behavior of the zero mode, as well as
the particle production per mode. Our results are that the expansion does {\em
not} prevent an epoch of preheating from occurring, but will in general {\em
reduce} the number of particles produced per physical volume compared to the
flat spacetime results. Furthermore, the energy lost from the zero mode 
due to redshift will typically prevent some of the unexpected behavior
found for the 
zero mode in Minkowski space\cite{us1,us2} from occurring in an
expanding universe. 

\section{The Formalism and the Model}

The correct way to describe the evolution of quantum fields in non-equilibrium
situations is the CTP formalism\cite{ctp} which describes the time evolution of
the density matrix of field theory. Details of the applications to field theory
are given in ref.\cite{us1,us2}.

A controllable yet non-perturbative approximation scheme is the large $N$
approximation of the $O(N)$ vector model, which has been generalized for
non-equilibrium situations\cite{largeN}. The lowest order term in
this approximation corresponds to mean field theory, but, at least in
principle, the next to leading terms that would include effects such as
scattering, could be computed. Such corrections will be important in any
discussion of how the particles produced during preheating actually thermalize,
and in particular for any estimates of the reheating temperature.

The $O(N)$ model in an FRW universe can be formulated in the following
manner\cite{usfrw}. The Lagrangian density is:

\begin{eqnarray}
{\cal{L}} &=& a^3(t)
\left[\frac{1}{2}\dot{\phi}^2(\vec{x},t)-\frac{1}{2}
\frac{(\vec{\nabla}\Phi(\vec{x},t))^2}{a(t)^2}-V(\vec{\phi}\cdot
\vec{\phi})\right] \nonumber \\
V(\sigma, \vec{\pi} ) &=& -\frac{1}{2} m^2 \vec{\phi} \cdot \vec{\phi} + \frac{
\lambda}{8N} ( \vec{\phi} \cdot \vec{\phi} )^2 \label{onLagrangian}
\end{eqnarray}
for fixed $\lambda$ in the large $N$ limit and $m^2>0$ (symmetry
breaking). Here $a(t)$ is the FRW scale factor, and $\vec{\phi} = (\sigma,
\vec{\pi} )$ is an $O(N)$ vector where $\vec{\pi}$ represents the $N-1$
``pions''. For simplicity, we will only consider the case of flat spatial
sections, and we will set the {\em renormalized} coupling to the Ricci scalar
$R$ to zero.

We decompose the $\sigma$ field as
\begin{equation}
\sigma  (\vec{x},t ) = \sigma_0(t)+ \chi ( \vec{x},t).
\end{equation}
The large $N$ limit can be implemented consistently with the following
Hartree-like factorization (for more details the reader is referred to
ref.\cite{us2}):
\begin{equation} 
\vec{\pi}(\vec x, t) = \psi(\vec x, t)
\overbrace{\left(1,1,\cdots,1\right)}^{N-1} \; \; , \; \; \sigma_0(t) = \phi(t)
\sqrt{N}, \label{filargeN}
\end{equation}
with
\begin{equation}
\langle \psi^2 \rangle \approx {\cal{O}} (1) \; , \; \langle \chi^2 \rangle
\approx {\cal{O}} (1) \; , \; \phi \approx {\cal{O}} (1).
\label{order1}
\end{equation}
The leading contribution is obtained by neglecting the $ {\cal{O}} ({1}\slash
{N})$ terms in the formal large $N$ limit.

If $V_k(t)$ are the mode functions (for comoving wavenumbers $k$) corresponding
to $\psi(\vec{x}, t)$, then the evolution equations for $\phi(t)$ and the
$V_k(t)$ are:
\begin{eqnarray}
\ddot{\phi}(t) + 3 H \dot{\phi}(t) + \phi (t) \left[ m^2 +\frac{\lambda}{2}
\phi^2 (t) + \frac{\lambda}{2} \langle \psi^2 (t) \rangle \right] & =& 0
\label{zeromodeeqn} \\ \left[ \frac{d^2}{d t^2} +3 H \frac{d}{dt}+
\frac{k^2}{a^2(t)} + m^2 + \frac{\lambda}{2} \phi^2 (t) +
\frac{\lambda}{2}\langle \psi^2 (t) \rangle \right] V_{k} (t)& = & 0,
\end{eqnarray}
\begin{eqnarray}
\langle \psi^2 ( t ) \rangle & = & \int \frac{d^3 k}{( 2 \pi)^3}
\frac{\left| V_k 
(t) \right|^2 }{2W_{k}} \label{fluctuation} \\ {W_{ k}} & = &
\sqrt{\frac{k^2}{a^2(t_0)}+m^2+\frac{\lambda}{2} \phi^2 (t_0)},
\label{initialfreqs} 
\end{eqnarray}
where $H \equiv \dot{a}(t)\slash a(t)$ is the Hubble parameter. The initial
conditions are $V_{k}(t_0)=1\ , \dot{V_{k}}(t_0)= -iW_{k}$.

After renormalization\cite{usfrw,us2}, we have:
\begin{eqnarray}
\eta''(\tau) + 3 \frac{a'(\tau)}{a(\tau)} \eta'(\tau) + {\cal
M}^2(\tau)\eta(\tau)  
& = & 0, \label{etanum} \\
\left[\frac{d^2}{d\tau^2} + 3 \frac{a'(\tau)}{a(\tau)} \frac{d}{d\tau}
+\frac{q^2}{a^2(\tau)} +{\cal M}^2(\tau)\right]
V_q(\tau) & = & 0, \label{Uknum}
\end{eqnarray}
\begin{equation}
{\cal M}^2(\tau) =  -1 + \eta^2(\tau) + g\Sigma(\tau),
\label{masssqnum}
\end{equation}
where $\tau \equiv mt$, $q \equiv k/m$, $\eta^2(\tau) \equiv \lambda \phi^2(t)/
2m^2$, and $g \equiv \lambda/8\pi^2$. Primes denote derivatives with respect to
$\tau$. The fluctuations operator $g\Sigma(\tau)$, which accounts for the back
reaction of the produced particles on the modes, is:
\begin{equation}
g\Sigma(\tau) = g \int^{\infty} dq \; q^2 \left\{\frac{|V_q(\tau)|^2}{W_q}
- \frac{1}{q a^2(\tau)} + \frac{\theta(q-\kappa)}{2q^3}\left[{\cal M}^2(\tau) 
- \frac16R(\tau) - \frac{a'^2(\tau_0)}{a^2(\tau)}\right]\right\},
\label{gSigma}  
\end{equation}
where $\kappa$ is a renormalization scale, $R(\tau)$ is the Ricci scalar in
units of $|m^2|$ and $V_q(\tau_0) = 1\ , V'_q(\tau_0)=-iW_q$.

We consider: (i) {\em slow roll} initial conditions
with $\eta(\tau_0)\simeq 0$. In this case some of the modes display spinodal
instabilities which then drive the production of particles; (ii) {\em chaotic}
initial conditions, $\eta(\tau_0)\gg 1$. The particle production in this case
is due to parametric amplification of the non-zero modes due to the evolution
of the zero mode. In both these cases, we will take as our background a
radiation dominated FRW universe, with scale factor given by $a(t) =
\left((t+t_0)\slash t_0\right)^{1/2}$. This analysis can be repeated
for a matter dominated universe with similar results, while the analysis of the
de Sitter case is quite different\cite{usDeS}.

\subsection{Slow-Roll Initial Conditions}

We take $\eta(\tau_0)$, to be very near the origin: $\eta(\tau_0)=10^{-5}\; ,
\; \eta'(\tau_0)=0$. The coupling is fixed at $g = 10^{-12}$ as would befit an
inflationary theory. Larger couplings will tend to speed up the dynamics, but
will leave the qualitative aspects unchanged.

\begin{figure}
\epsfig{file=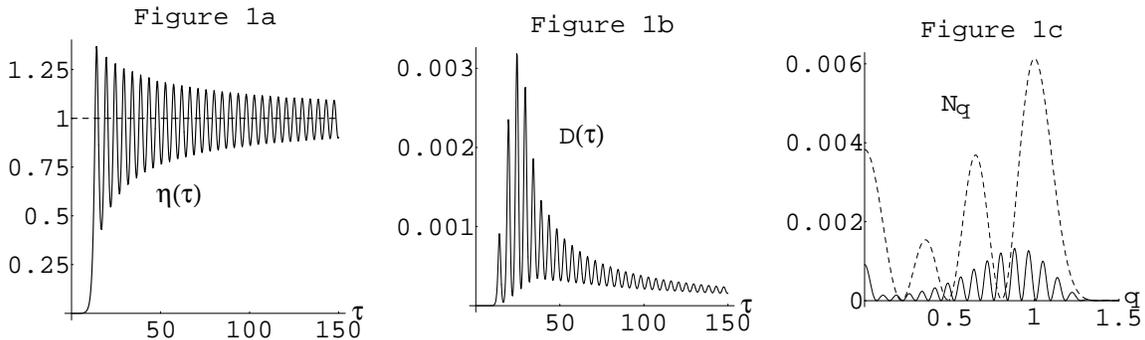}
\caption{(a) The zero mode $\eta(\tau)$ vs. $\tau$,
(b) the quantum fluctuation operator $D(\tau)$ vs. $\tau$,
and (c) the number of pions $N_q$ vs $q$ at $\tau = 30$ (dashed line),
$\tau=150$ (solid line) for the parameter values $\eta_0 = 10^{-5}, \eta'_0=0,
g = 10^{-12}, h_0 = 0.1$. \label{fig1}}
\end{figure}

Fig.(\ref{fig1}a) shows the behavior of $\eta(\tau)$ for an initial
(dimensionless) Hubble parameter $h_0 = 0.1$. We see that it makes one
excursion out to near the classical turning point and then tries to return to
its initial value. However, by this time, the expansion of the universe has
redshifted enough energy from the zero mode that it cannot come back to the
origin and begins to execute damped oscillations about the tree-level minimum
at $\eta=1$.

Fig.(\ref{fig1}b) shows the quantum fluctuations operator $D(\tau) \equiv
g\Sigma(\tau)$ during the evolution. We see that $D(\tau)$ grows due to
spinodal instabilities as $\eta(\tau)$ makes its first excursion; this is when
particle production occurs. However, at late times $D(\tau)$ tends towards
zero.

Fig.(\ref{fig1}c) shows the number of ``pions'' per unit physical volume, in
units of $1\slash g$, produced as a function of comoving dimensionless
wavenumber $q$ at various times. An important feature to notice here is that
this distribution has support at $q=0$.

In Minkowski space where $h_0=0$\cite{us2}, the zero mode finds a stable
minimum near the origin, while the fluctuations grow to $D(\tau\rightarrow
\infty)\rightarrow 1$. Furthermore, the momentum distribution of the produced
pions has essentially no support near $q=0$ in this situation. Thus the
addition of expansion induces a qualitative change in the dynamics of the
system and the power spectrum of fluctuations. 

Increasing $h_0$ yields essentially the same features, the only differences
being in the time scale involved. Thus for larger $h_0$, $\eta$ begins to
oscillate about the tree-level minima sooner, and the fluctuations begin their
decrease towards zero sooner. 

\subsection{Chaotic Initial Conditions}

\begin{figure}
\epsfig{file=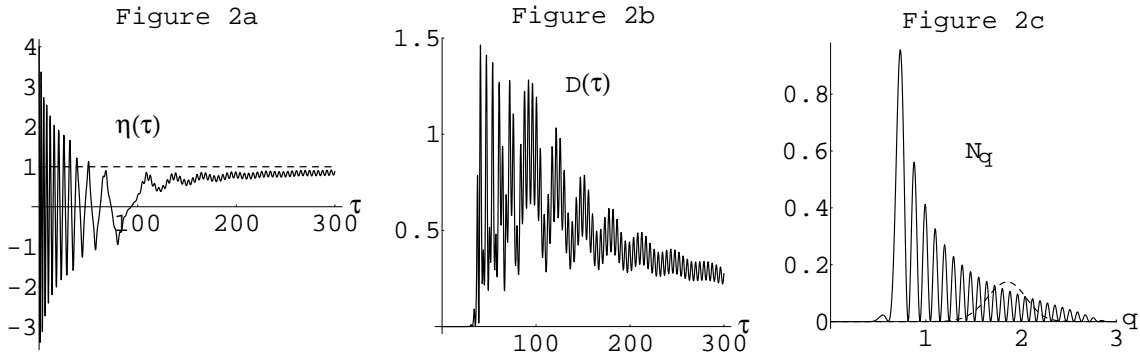}
\caption{(a) The zero mode $\eta(\tau)$ vs $\tau$,
(b) the quantum fluctuation operator $D(\tau)$ vs. $\tau$,
and (c) the number of pions $N_q$ vs $q$ at $\tau = 40$ (dashed line),
$\tau=200$ (solid line) for the parameter values $\eta_0 = 4, \eta'_0=0,
g = 10^{-12}, h_0 = 0.1$. \label{fig2}}
\end{figure}

We take $\eta(\tau_0)=4\ , \eta'(\tau_0)=0\ , h_0 = 0.1$ as our chaotic initial
conditions, again with $g = 10^{-12}$. In this case, the zero
mode has enough energy to be able to sample both minima for a period of time
until enough energy is lost both due to the redshifting as well as to particle
production to restrict it to small oscillations about one of the tree level
minima, in this case the one at $\eta = 1$ (Fig.(\ref{fig2}a)). This
behaviour is different than the Minkowski case \cite{par} where the zero
mode keeps oscillating over the two minima showing that the symmetry
is unbroken if the initial energy is larger than $V(\eta=0)$.

As in the slow-roll case, the fluctuations are again driven to zero at late
times as the zero mode reaches its minimum (Fig.(\ref{fig2}b)). However, the
behavior of the momentum distribution of particles (Fig.(\ref{fig2}c)) is quite
different, looking much more like the Minkowski slow-roll space distribution
than the slow-roll one above.

Changing the initial value of $\eta(\tau_0)$ will change the number of times
both ground states are sampled; for $\eta(\tau_0)>4$, we have more oscillations
back and forth before enough energy is drained away to make the zero mode
settle into one of the vacua. If the initial Hubble parameter is increased,
for fixed $\eta(\tau_0)$, the energy of the zero mode is redshifted away
faster, so that fewer large amplitude oscillations are executed and fewer
particles per {\em physical} volume are produced.

\section{Late Time Analysis}

An interesting by-product of our numerical analysis concerns the late time
behavior of the field. The damping term in the zero mode equation suggests that
there will be stationary solutions at late times. Such solutions must obey the
large N Ward identity
\begin{equation}
\eta_{\infty}(-1+\eta_{\infty}^2 + g \Sigma_{\infty}) =0,\label{sumrule}
\end{equation}
where the subscript $\infty$ denotes the late time value of the relevant
quantity.  Our numerical results confirm that eq.(\ref{sumrule}) is satisfied
to one part in $10^{10}$. The same sum rule was found to be satisfied in the
Minkowski case as well. For $\eta_{\infty} \neq 0$ the term inside the bracket
must vanish, leading to a mode equation for massless Goldstone bosons, which
can be solved exactly in a radiation dominated universe:
\begin{equation}
V_q (\tau) = \frac{1}{a(\tau)}\left[c_q \;
\exp\left(2iq\tau_0^{1/2}(\tau+\tau_0)^{1/2}\right) + d_q \;
\exp\left(-2iq\tau_0^{1/2}(\tau+\tau_0)^{1/2}\right)\right],
\end{equation}
where $c_q$ and $d_q$ are determined by the initial conditions and the
dynamical evolution. For $q \neq 0$ and late times, $\tau \gg 1/(q^2 \tau_0)$,
the phases average out so that $|V_q(t)|^2 \sim a(\tau)^{-2}$ and the
fluctuation contribution is driven to
\begin{equation} \label{fluctlt}
g\Sigma(\tau)\simeq \frac{{\cal C}}{\tau+\tau_0}\propto a(\tau)^{-2},
\end{equation}
where ${\cal C}$ is a constant which depends on the dynamics of the evolution
and initial conditions. Using eq.(\ref{sumrule}), the zero mode then behaves
as:
\begin{equation} \label{zeromdlt}
\eta^2(\tau) \simeq \frac{\tau+\tau_0-{\cal C}}{\tau+\tau_0}.
\end{equation}

If $\eta(\tau_0) \neq 0$, we find that $V_0(\tau)$ tends to a constant at late
times. This explains the appearance of the peak in $N_q$ at $q=0$. Since
$\eta(\tau_0) = 0\ , \eta'(\tau_0)=0$ is a fixed point of the zero mode
equation, eq.(\ref{sumrule}) is satisfied with $g\Sigma_{\infty}=1$. In this
case only the $q \rightarrow 0$ modes contribute substantially to the
fluctuations and the power spectrum of the fluctuations becomes both peaked at
zero momentum and narrower at longer times. This analysis reproduces our
numerical results to high accuracy.

There is a striking contrast between the radiation dominated and the de Sitter
cases. In the latter, for $\eta(\tau_0) \neq 0$, there are consistent
stationary solutions for which $\eta_{\infty}$ is not a minimum of the tree level
potential and the fluctuation contribution $g\Sigma_{\infty}$ is driven to a
nonzero constant\cite{usDeS}.

\section{Conclusions}

We have generalized our previous results on the quantum dynamics of
the $O(N)$ model in Minkowski space to FRW universes. In particular, we find
that preheating can still occur in an expanding universe, though the actual
amount of particle production depends sensitively on the size of the initial
Hubble parameter relative to the mass of the field. 

We also showed that the late time dynamics is determined by the
sum rule in eq.(\ref{sumrule}). This, together with the behavior of the modes,
drives the system to satisfy the sum rule by damping the fluctuations to zero
and allowing the zero mode to find one of the tree level minima. 
This indicates that the symmetry is broken at late times\cite{par}. 
We see no evidence of symmetry restoration\cite{symrest}.

The next step in this program is clearly to attempt to allow the scale factor
to evolve dynamically and understand how the back reaction of the produced
particles influences the dynamics of $a(t)$. We are currently setting up the
formalism to tackle this problem.

\acknowledgements 

D.B. thanks the N.S.F. for support under grant awards: PHY-9302534 and
INT-9216755. R. H. and D. C. were supported by DOE grant
DE-FG02-91-ER40682. A.S and M.S were supported by the N.S.F under grant
PHY-91-16964.

\end{document}